\documentclass[slac_one]{revtex4}
\usepackage{graphicx}
\usepackage{fancyhdr}
\begin{document}

\title{{\small{2005 International Linear Collider Workshop - Stanford,
U.S.A.}}\\ 
\vspace{12pt}
The 2mrad horizontal crossing angle IR layout for a TeV ILC} 

%

\author{R.~Appleby\footnote{On behalf of the SLAC-BNL-UK-France task force, r.b.appleby@dl.ac.uk}, D.~Angal-Kalinin}
\affiliation{ASTeC, Daresbury Laboratory, Warrington, WA4 4AD, England}

\author{P.~Bambade, B.~Mouton}
\affiliation{LAL, Campus Universitaire, F-91898 ORSAY CEDEX, France}

\author{O.~Napoly, J.~Payet}
\affiliation{DAPNIA-SACM, CEA Saclay, 91191 Gif/Yvette CEDEX, France}

\author{A.~Seryi, Y.~Nosochkov}
\affiliation{Stanford Linear Accelerator Center, Menlo Park, CA, USA}

\begin{abstract}
The current status of the 2mrad crossing angle layout for the ILC is reviewed. The scheme
developed in the UK and France is described and the performance discussed for a TeV machine. Secondly, the
scheme developed at SLAC and BNL is then studied and modified for a TeV machine. We find that both 
schemes can handle the higher energy beam with modifications, and share many common features.
\end{abstract}

\maketitle

\thispagestyle{fancy}


\section{INTRODUCTION}
\label{intro}

In this article, we describe the recent development of the~2mrad horizontal crossing angle 
scheme for the ILC. To date, two parallel designs have emerged: the first coming from the UK
and France and the second from SLAC and BNL. We shall describe both schemes here, although
they share many common features and are now being developed in parallel under the
unified SLAC-BNL-UK-France task force collaboration. The work presented focuses on the performance
at~1~TeV. The benefits of the scheme are well documented~\cite{Appleby:2004df}: for small crossing angles, the loss of 
luminosity is small (crab correction may not be necessary and it may be possible to partially correct this loss by 
exploiting the finite~$\eta'$ at the IP for local chromaticity correction lattices), no electrostatic separators 
or kicker magnets are needed and the conditions are improved for physics (e.g. better forward coverage). 
A known weakness of this scheme is however its more difficult conditions for extracting cleanly the spent 
disrupted and energy-degraded beam, in comparison with larger crossing-angle schemes where separate
magnetic channels can be used for in-and outgoing beams.

The work presented here covers the designs developed at SLAC, BNL, the UK and France. In section~\ref{euroscheme} we shall
describe the scheme developed in Europe and discuss its performance at~1~TeV. In section~\ref{slacscheme}, we shall
discuss the performance of the SLAC/BNL scheme, when extended to~1~TeV from the initial design at~500~GeV, and 
we shall draw our conclusions in section~\ref{conc}.

\section{The UK/FRANCE EXTRACTION LINE DESIGN AND PERFORMANCE AT 1 TEV}
\label{euroscheme}

In this section we shall describe the 2mrad interaction region layout and extraction line
for the 2mrad horizontal crossing angle scheme.

The final doublet magnets have been optimised for the extraction of a~500~GeV beam. A similar optimisation 
exists for the baseline beam energy of~250~GeV. It has been shown that the doublet parameters calculated 
for a high energy beam also provide acceptable extraction at lower energy. The superconducting magnet
closest to the IP, denoted QD, is of critical importance to the IR layout properties and is chosen to be a
LHC low-$\beta$-insertion quadrupole. This provides the required aperture and field strength to accommodate 
both the incoming and outgoing (disrupted) beams. Note that the outgoing beam possesses a long low energy tail 
and, by virtue of
the crossing angle, is off-axis in QD. The other final doublet magnet, QF, is constructed from a normal conducting
magnet and is separated from QD by~3m. For the TeV machine, QD and QF are~2.3m and~1.9m long, respectively.
After passing through QD, the outgoing disrupted beam enters the extraction line, which provides beam 
transport to the dump and downstream diagnostics (the geometry is fixed by the 
linear matrix element~$R_{22}$ from the IP to the exit of QD).

The LHC low-$\beta$-region quadrupoles are constructed 
from~NbTi and can achieve a gradient of~215~Tm$^{-1}$ with an available aperture for the beam 
of~62mm. Note that higher gradients are
currently under development, which will aid the present application. LHC studies of the tolerable power depostion 
indicate local and integral values of~0.4~mWg$^{-1}$ and~5~Wm$^{-1}$ respectively; this approximately
translates into a maximum power deposition from charged particles into QD of around~10W.

Note that in all these studies, unless otherwise noted, we assume the parameters of the TeV ILC parameters working
group and, where possible, assume the worst possible parameter set for extraction. 
In this work, we follow~\cite{applebyandbambade} and assume a photon cone half-opening
angle of~0.5mrad in all cases. This ensures that the cone contains all the photon power, apart from~100W. This remaining 
power needs to be lost in suitable collimators. Ensuring 
extraction of the photons past QF immediately requires a crossing angle of greater than~1.6mrad.

\begin{figure}
\begin{center}
\begin{minipage}{0.49\textwidth}
\includegraphics[width=6cm]{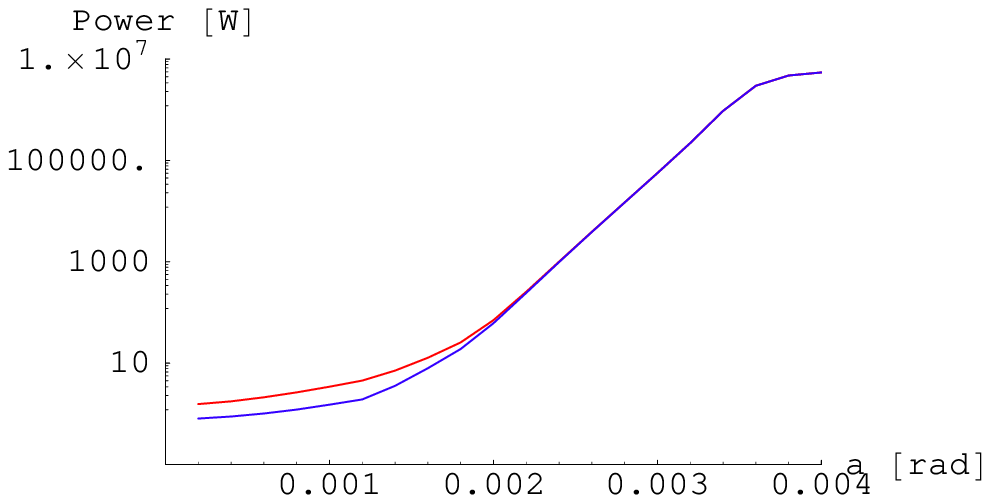}
\end{minipage}
\begin{minipage}{0.49\textwidth}
\includegraphics[width=6cm]{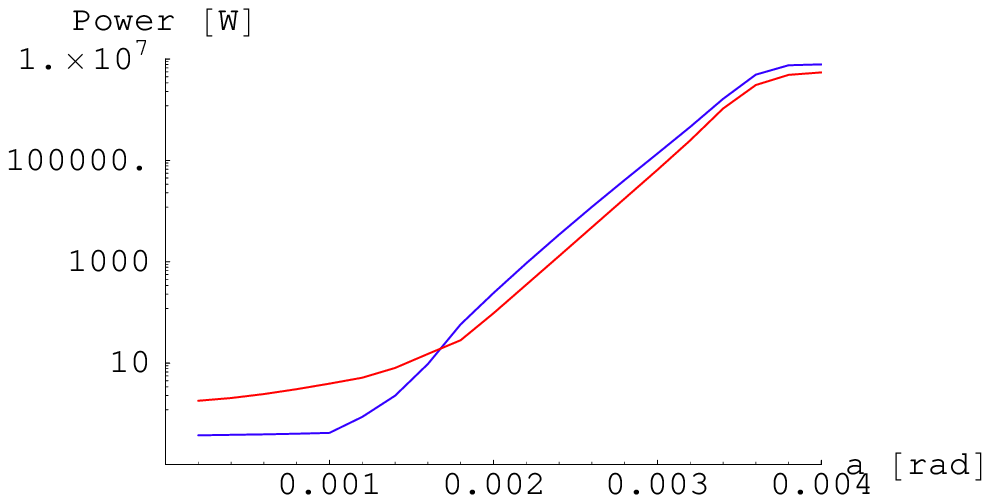}
\end{minipage}
\caption{The power losses into QD, as a function of crossing angle. The left-hand plot is for the case of 
a TeV machine with~2$\times$10$^{10}$ bunch population and the right-hand plot is a TeV machine with the 
vertical offset of~150nm at the IP. In both cases the blue line (the lower curve at small angle) denotes
the power loss with the beam size effect turned on and the red line (the upper curve at small angle) 
denotes the power loss with the beam size effect turned off. Note that for the right-hand plot, both
curves have the vertical offset.}
\label{figqdel}
\end{center}
\end{figure}

Figure~\ref{figqdel} shows the power deposition into QD from charged particles produced during the beam-beam
interaction, as a function of crossing angle. These power deposition calculations are described in detail
in~\cite{applebyandbambade2}. The charged particle loss comes from two sources: the low energy tail of the disrupted
beam and from radiative Bhabha events produced in the interaction of colliding particles (also refered to as 
the ``Compton tail''). The latter contribution is suppressed in regions of phase space of low transverse momentum 
exchange, where the virtual photons participating in the scattering process can have a transverse position 
indeterminacy exceeding the transverse beam size. The suppression from this so-called beam-size effect is 
illustrated in the curves of fig. 1. Conservatively, the results without it are however used for the worst-case 
scenarios considered here. If we assume a maximun power loss of~10W, we find that for
the case of a TeV machine with a bunch population of~2$\times$10$^{10}$, we can tolerate a crossing angle
no larger than~1.6mrad. This result is dominated by the Compton tail. The other case we have considered
in figure~\ref{figqdel} shows a larger permitted crossing angle, and hence easier extraction. 
This case, with a vertical offset of~150nm at the IP, is studied because it maximises the low energy tail of the 
disrupted beam. For further details of
these calculations see~\cite{applebyandbambade2}. 
All of these curves were produced using the US cold machine parameters, for
which the key parameters (for this study) are similar to the WG 1 new ILC nominal parameters~\cite{wg1}.

\begin{figure}
\begin{center}
\begin{minipage}{0.49\textwidth}
\includegraphics[angle=-90,width=6.5cm]{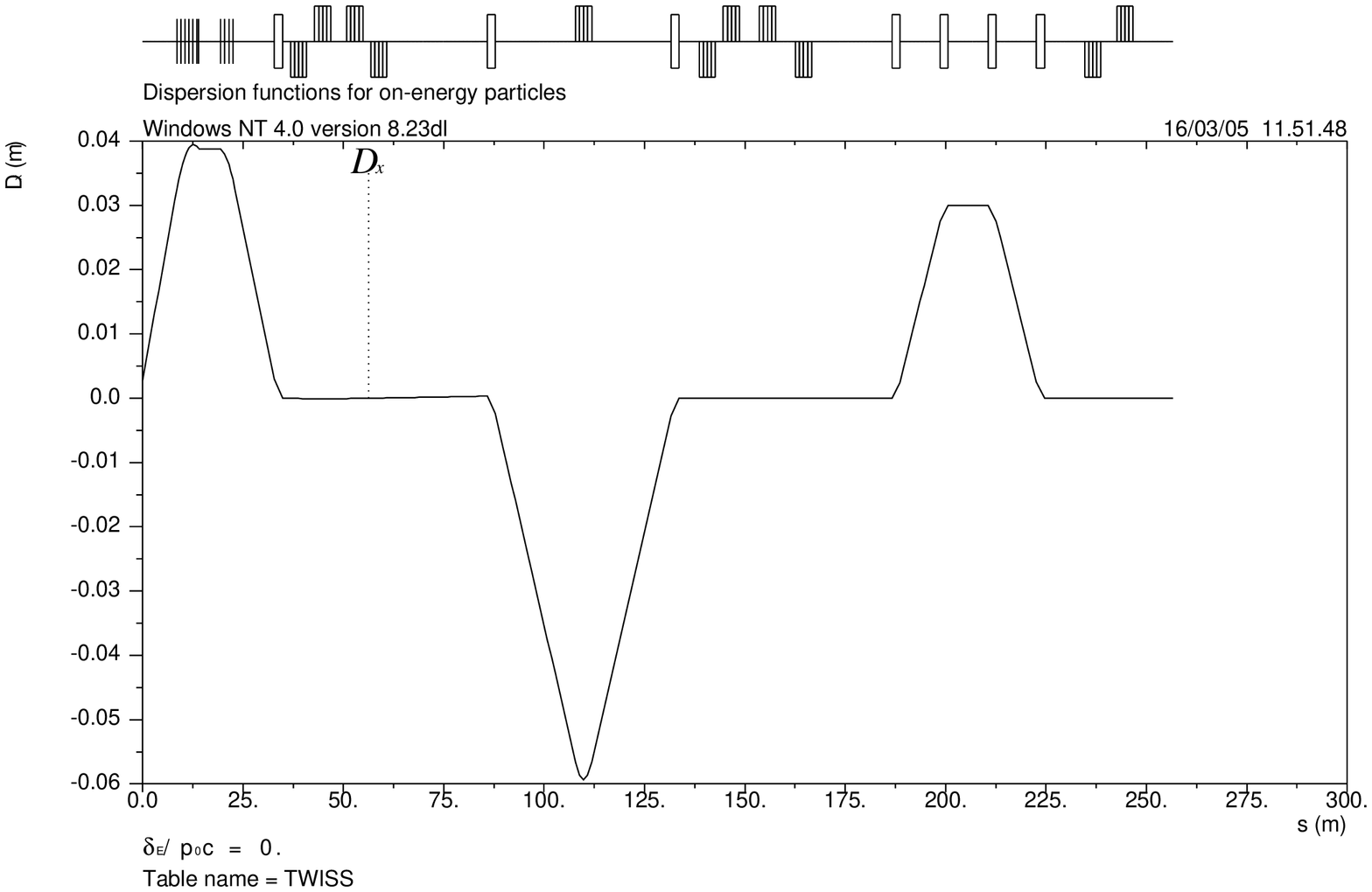}
\end{minipage}
\hfill
\begin{minipage}{0.49\textwidth}
\includegraphics[width=6cm]{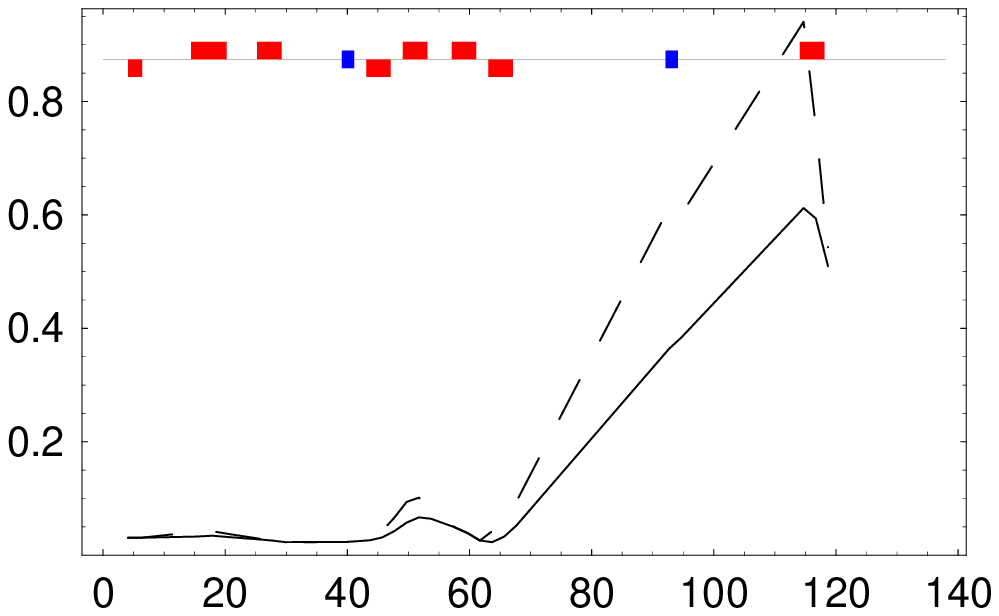}
\end{minipage}
\caption{The linear dispersion (left) and power loss curve (right) for the 1 TeV extraction line.  
The right-hand plot show the required magnet apertures ([m]) to achieve~0.5ppm beam loss at each element, 
for the case of the TeV machine nominal parameters (with (dashed line) and without (solid line) a vertical offset).}
\label{figplots}
\end{center}
\end{figure}

The purpose of the extraction line is to transport the disrupted beam to the dump with controlled 
and acceptable losses, handle the undisrupted beam in the case of single beam operation (e.g. during 
commissioning) and to
provide diagnostics on the disrupted beam. The possible diagnostics include Compton polarimetry and 
energy spectrometry. The linear dispersion  
for the extraction line is shown in the left-hand plot of figure~\ref{figplots}. 
The linear dispersion allows identification of the major optical structures, denoted by regions of non-zero 
dispersion. The first region is the dispersion generated by the beam being off-axis in QD (the first
magnet in the line). This dispersion is cancelled by a mirroring dipole. The second dispersive
region is a ``bend-back'' linear-achromatic structure, which bends the beam parallel to 
the beam at the IP. This is necessary to perform Compton polarimeter and ``undo'' the spin precession
occuring in upstream dipoles. The third region is the diagnostic chicane; the current requirements
for this chicane is a secondary beam focus and some dispersion, such that the dispersive beam size
dominates the betatron beam size. Requirements of beamsize/laser wavelength correlation and of
minimum background requirements are under study. 
The first magnet in the extraction line is of critical importance, due to the need to place the first
disrupted beam magnet as close as possible to the IP, balanced against the constraints of transverse space. For
this reason, a half-quadrupole current sheet magnet is used~\cite{teslatdr}. 
These magnets, developed for the TESLA extraction
line~\cite{teslatdr} provide the required aperture and maximum pole tip field whilst allowing the 
magnet to be placed~8m after
the exit of QD (with~50mm transverse separation). Note that other magnet designs are under study.

The right-hand plot of figure~\ref{figplots} shows the optimisation of the apertures for a fixed power loss. 
We have used the TeV machine nominal parameters and demanded a power loss of~0.5ppm at each element. 
We also considered a 
vertical offset of~120nm. The apertures were calculated using explicit all-orders ray-tracing of the disrupted beam. 
The calculated apertures are acceptable, although the very downstream apertures for the TeV machine need
some optimisation. 

The study of this scheme is on-going. It is necessary to include the effects on the outgoing beam of all the 
fields from nearby magnets used to focus the incoming beam, in particular QF. This can be done by way of 
suitable multipole expansions, once the detailed field map is computed. It has been shown to be particularly 
critical to predict reliably the transport of the low-energy tail of the extracted beam~\cite{yuritalk}. 
Furthermore the magnet technology is of prime importance, and substitutes for the half-quadrupole described here are under 
consideration, along with the use of sextupoles in the extraction line to aid the control of the low energy beam tail. 
The layout of the charged particle and photon beam dumps is also a topic for further work, along with
detailed calculations of backgrounds in the detector and diagnostic sections. 

\section{The SLAC-DEVELOPED DESIGN EXTENDED TO 1 TEV}
\label{slacscheme}

\begin{figure}
\begin{center}
\begin{minipage}{0.49\textwidth}
\includegraphics[width=5.5cm]{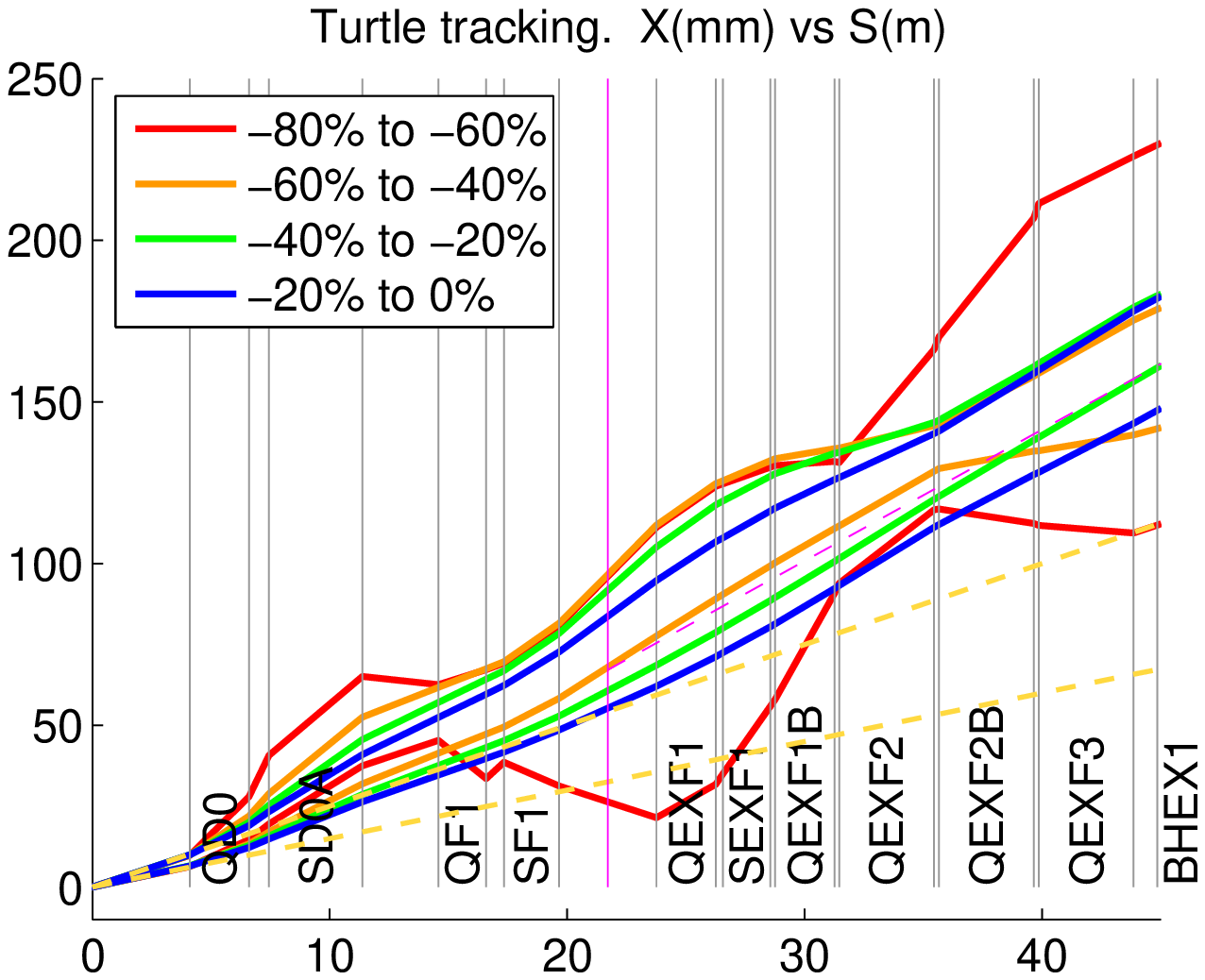}
\end{minipage}
\hfill
\begin{minipage}{0.49\textwidth}
\includegraphics[width=5.5cm]{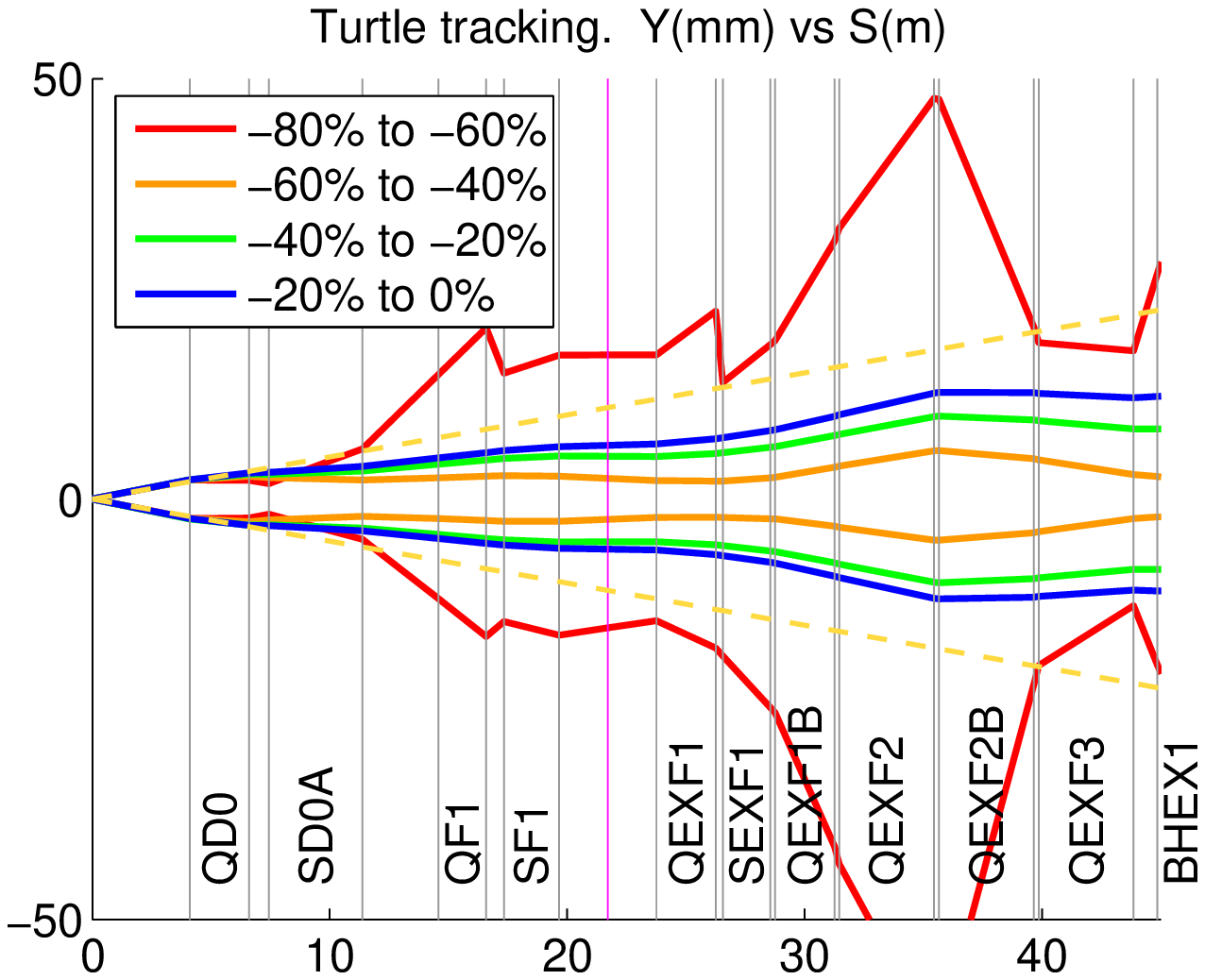}
\end{minipage}
\caption{The tracking of low energy tail particles along the SLAC-designed 2mrad extraction line, for a 1 TeV machine.
The left-hand plot shows the horizontal beam envelope and the right-hand plot shows the vertical beam envelope.}
\label{figslac}
\end{center}
\end{figure}

In this section, we describe the performance of the SLAC/BNL-designed~2~mrad scheme, when extended to the upgrade 
centre-of-mass energy of~1~TeV.
This scheme, described in~\cite{yuritalk}, was studied at~1~TeV using the 
nominal parameters and a vertical offset of~100nm. This gives a low energy Beamstrahlung tail reaching down to~18\% of 
maximum energy, when computed with high statistics.  
Extending the design to~1~TeV required a number of modifications to the initial 
design for~500~GeV: proper scaling 
of the lengths of sextupoles to keep the pole tips within maximum achievable fields, increasing~L* from~3.51m to~4.1m and
reducing the length of~QD0 to reduce over focusing of low energy tail particles. In order to 
reduce the losses of the low energy tail particles on the magnets, a collimator was introduced in the vertical 
plane within the final doublet.  The horizontal and vertical beam envelopes for this scheme are shown 
in figure~\ref{figslac}, for particles in different energy ranges. 
Note that this is a first look at the TeV upgrade of this scheme and 
further optimisation and power losses on the magnets needs to be evaluated. 

\section{CONCLUSION}
\label{conc}
We have described the 2mrad horizontal crossing angle schemes under development for the ILC. The first part described
the scheme developed in the UK and France, and focused on its propeties at~1~TeV. The second describes the
study of the SLAC and BNL scheme from these proceedings, when extending the design to~1~TeV.  Both schemes at 1 TeV can 
work with nominal parameters but the high luminosity case seems difficult. Further development work is ongoing. Finally
we note that both schemes share many common features.

\begin{acknowledgments}

We would like to thank Cherrill Spencer and Brett Parker for useful conversations and advice.

\end{acknowledgments}

\end{document}